\pdfoutput=1
\documentclass[aps,twocolumn,amsmath,amssymb,preprintnumbers,superscriptaddress,floatfix]{revtex4-1}
\usepackage[utf8]{inputenc}
\usepackage{newtxtext}
\usepackage[upint]{newtxmath}
\usepackage{microtype}
\usepackage{textcomp}
\usepackage{eucal}
\usepackage{bm}

\usepackage{graphicx}
\usepackage{enumerate}
\usepackage{amsfonts}
\usepackage{amsmath}
\usepackage{amssymb}
\usepackage{siunitx}
\usepackage{color}
\usepackage{soul}
\usepackage[colorlinks,allcolors=blue]{hyperref}
\usepackage[capitalize]{cleveref}

\let\phi=\varphi
\let\epsilon=\varepsilon

\newcommand{\B}[1]{\bm{#1}}

\newcommand{\V}[1]{\check{#1}}
\renewcommand{\H}[1]{\hat{#1}}

\newcommand{\eg}{\emph{e.g.\ }}
\newcommand{\ie}{\emph{i.e.\ }}

\definecolor{DarkRed}{rgb}{0.80,0,0}
\definecolor{DarkBlue}{rgb}{0,0,0.80}
\definecolor{Green}{rgb}{0,0.8,0}
\definecolor{Purple}{rgb}{0.55,0,0.55}
\definecolor{Orange}{rgb}{1,0.6,0}
\definecolor{Gray}{rgb}{0.75,0.75,0.75}

\newcommand{\Tc}{T_{\textsc{c}}}
\newcommand{\Isc}{I_{\textsc{sc}}}
\newcommand{\Voc}{\Delta V_{\textsc{oc}}}
\newcommand{\Gphi}{G_\varphi}
\newcommand{\GD}{G_{\textsc{n}}}
\newcommand{\GT}{G_{\textsc{t}}}

\newcommand{\gr}{\H{g}^{\textsc{r}}}
\newcommand{\ga}{\H{g}^{\textsc{a}}}
\newcommand{\gk}{\H{g}^{\textsc{k}}}
\newcommand{\Ir}{\H{\B{I}}^{\textsc{r}}}
\newcommand{\Ia}{\H{\B{I}}^{\textsc{a}}}
\newcommand{\Ik}{\H{\B{I}}^{\textsc{k}}}

\begin{document}
\title{Complete magnetic control over the superconducting thermoelectric effect}
\author{Jabir Ali Ouassou}
\affiliation{Center for Quantum Spintronics, Department of Physics, Norwegian University of Science and Technology, NO-7491 Trondheim, Norway.}
\author{César González-Ruano}
\affiliation{Departamento Física de la Materia Condensada C-III, INC and IFIMAC, Universidad Autónoma de Madrid, Madrid 28049, Spain.}
\author{Diego Caso}
\affiliation{Departamento Física de la Materia Condensada C-III, INC and IFIMAC, Universidad Autónoma de Madrid, Madrid 28049, Spain.}
\author{Farkhad G. Aliev}
\affiliation{Departamento Física de la Materia Condensada C-III, INC and IFIMAC, Universidad Autónoma de Madrid, Madrid 28049, Spain.}
\author{Jacob Linder}
\affiliation{Center for Quantum Spintronics, Department of Physics, Norwegian University of Science and Technology, NO-7491 Trondheim, Norway.}
\begin{abstract}
	Giant thermoelectric effects are known to arise at the interface between superconductors and strongly polarized ferromagnets, enabling the construction of efficient thermoelectric generators.
	We predict that the thermopower of such a generator can be completely controlled by a magnetic input signal:
	not only can the thermopower be toggled on and off by rotating a magnet, but it can even be entirely reversed.
	This \emph{in situ} control diverges from conventional thermoelectrics, where the thermopower is usually fixed by the device design.
\end{abstract}
\maketitle

\section{Introduction}
It has been known for two centuries that \emph{thermo\-electric effects} can arise at the interface between two different metals or semiconductors~\cite{Rowe.1995}.
Such junctions can be used to convert heat flows into electricity (Seebeck effect), or to transport heat via electricity (Peltier effect).
Applications include \emph{thermoelectric generators} that convert waste heat into useful electric power, \emph{thermoelectric coolers} acting as purely electric heat pumps, and \emph{thermocouples} used in many digital thermometers.
Compared to other technologies such as heat engines and heat pumps, thermoelectric devices are compact, require little maintenance, and contain no moving parts or circulating fluids.
This makes them an ideal choice for \eg sensitive lab equipment and deep space missions.
However, conventional thermoelectrics suffer from low efficiencies which has prevented their adoption in other domains.
Therefore, a significant research interest lies in identifying better material platforms for thermoelectric devices.
In this paper, we focus on thermoelectric platforms that may be useful under low-temperature laboratory conditions.

\begin{figure}[b!]
	\includegraphics[width=\columnwidth]{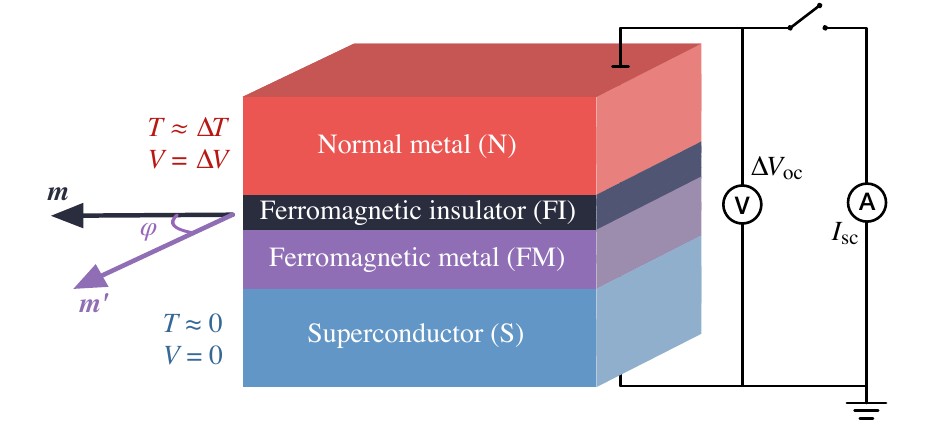}
	\caption{
		Proposed experimental setup.
		The junction's thermopower $S(\varphi)$ is determined by the angle~$\varphi$ between the magnetic orientations $\B{m}$ and $\B{m}'$ of the two ferromagnets.
		The latter is rotated using an external magnetic field, providing \emph{in situ} control over the thermopower.
		When the normal metal is heated using \eg a light-emitting diode, one can measure a magnetically controlled thermoelectric voltage $\Voc(\varphi)$ (open circuit) or current $\Isc(\varphi)$ (short circuit).
	}
	\label{fig:setup}
\end{figure}

One metric for comparing thermoelectric platforms is the \emph{thermopower} ${S \equiv -\Delta V/\Delta T}$ of a thermoelectric generator, where $\Delta T$ is the driving temperature difference and $\Delta V$ the generated electric~potential.
Recently, it has been shown theoretically \cite{Kalenkov.2012, Machon.2013, Ozaeta.2014, Kalenkov.2014, Machon.2014, Kalenkov.2015k8, Hwang.2016, Linder.2016eha, Bathen.2017, Rezaei.2018, Dutta.2020} and experimentally \cite{Kolenda.2016, Kolenda.2016fd8, Kolenda.2017, Heidrich.2019} that a \emph{giant thermoelectric effect}---with thermopowers up to $\sim\!\SI{100}{\upmu\volt\per\kelvin}$---can be realized by interfacing superconductors with ferromagnets.
For comparison, metals typically have thermopowers of \SIrange{1}{10}{$\upmu$\volt\per\kelvin} at room temperature, which vanishes at low temperatures~\cite{Rowe.1995}.
Potential applications were promptly proposed, including electron cooling~\cite{Rouco.2018}, electron thermometry~\cite{Giazotto.2015}, and radiation detectors~\cite{Heikkila.2018, Chakraborty.2018}.
These findings have stimulated further studies of thermoelectric effects in superconductors 
\cite{dutta_prb_17, marchegiani_prl_20, keidel_prr_20, savander_prr_20, bobkova_prb_21} and also had an impact on phase-coherent caloritronics~\cite{Fornieri.2017};
\eg in Josephson junctions the giant thermoelectric effect manifests as a \emph{thermophase}~\cite{Giazotto.2015z7}.
For an overview of nonequilibrium effects in super\-conductor/\-ferro\-magnet structures, see Ref.~\cite{Bergeret.2018}.

In this paper, we predict another key advantage of super\-conductor/ferro\-magnet hybrids as a thermoelectric platform:
\emph{The thermopower can be tuned from a large positive value to a large negative value by rotating or inverting an in-plane magnetic field.}
This effect is realized by coupling the superconductor to two noncollinear ferromagnets, where one magnet dominates the spin splitting of the superconducting density of states, while the other dominates the spin filtering of thermal excitations.
This is in contrast to most previous studies of the giant thermoelectric effect in superconducting hybrids, where the spin splitting and filtering have been along the same magnetic axis, and the focus has been on \emph{maximizing} rather than \emph{controlling} the thermopower.
We elaborate on the control mechanism and resulting physical predictions in \cref{sec:results}.

\cref{fig:setup} illustrates our proposal for a magnetically controlled thermoelectric generator.
A layered structure is constructed from a superconducting reservoir~(S), weak ferro\-magnetic metal~(FM), fully spin-polarized ferro\-magnetic insulator~(FI), and normal-metal reservoir~(N).
The S is grounded and cooled far below its critical temperature~$\Tc$, while the N is heated to a higher temperature~$\Delta T$ using \eg a light-emitting diode.
We predict that such a device will exhibit a giant thermoelectric effect that is highly sensitive to the magnetic misalignment~$\varphi$ between FM and~FI.
The resulting thermopower can \eg be antisymmetric $S(\varphi) \sim \cos\varphi$, asymmetric $S(\varphi) \sim 1 + \cos\varphi$, or symmetric $S(\varphi) \sim 1$, depending on the junction parameters.
Experimentally, this is measured as either an \emph{open-circuit voltage} $\Voc(\varphi)$ or \emph{short-circuit current} $\Isc(\varphi)$, depending on the state of the electric switch in \cref{fig:setup}.
During the experiment, the angle~$\varphi$ can \eg be controlled by rotating the sample in an externally applied magnetic field.
Here, we consider $\bm{m}$ and $\bm{m}'$ that are restricted to the thin-film plane, in which case flux injection from the applied field is negligible.

The proposed effect may also be useful for applications where \emph{in situ} control over the magnetic misalignment $\varphi$ is difficult.
For example, conventional Peltier elements are constructed from alternating pillars of $p$- and $n$-doped semiconductors, which are connected thermally in parallel and electrically in series.
These materials are chosen because they have comparatively high thermopowers with opposite signs.
This enables electricity to flow in opposite directions in neighboring pillars even though every pillar transports heat in the same direction.
Our results suggest that at low temperatures, S/FM/FI/N pillars can replace both $p$- and $n$-doped pillars; the ``doping'' of a given pillar is then determined by whether its $\bm{m}$ and $\bm{m}'$ are parallel or antiparallel. 

The ideal candidate heterostructure to verify the predictions presented in this work would be fully epitaxial superconductor-spin valve tunnel junctions, in which the schematic structure would be superconductor/soft ferromagnet/hard ferromagnet.
Epitaxy should help minimize magnetic textures in the ferromagnetic layers, which could otherwise induce vortices in the superconductor through dipolar fields.
Moreover, symmetry-dependent spin filtering, for example in Fe/MgO, may increase the effective spin polarization of the tunneling to around 0.7--0.8.
Such an interface would behave similarly to the ferromagnetic insulators discussed in our manuscript, since MgO is itself insulating and the Fe/MgO interface is significantly more polarized than pure Fe.
Additionally, the structures should ideally have lateral dimensions of a few tens of microns to minimize edge-related magnetic charges.
To establish and vary a finite temperature gradient over the structure, it is desirable to introduce thin insulating barriers between the superconducting and ferromagnetic electrodes.
A somewhat smaller barrier between the two ferromagnets composing the spin valve would provide a tunnelling magnetoresistance signal to precisely control the magnetic orientation of the soft ferromagnet (near the superconductor), while providing the conditions to apply the main temperature gradient over the stronger barrier between the soft ferromagnet and the superconductor (\ie the main thermal conduction bottleneck).
An optimal candidate to verify the predicted effects is the V/MgO/Fe/MgO/Fe/Co heterostructure, which is known to grow epitaxially and with a second MgO barrier (\ie between the soft and hard FMs) being about 4 times more transparent than the barrier between the superconductor and the soft Fe layer \cite{Martinez2018}.
Alternative candidates to investigate the predicted effects might be oxide-based devices used for epitaxial superconducting spintronics \cite{Visani2012}.
However, those materials are much more resistive, so the main temperature gradients would then drop in the electrodes themselves and not between the materials constituting the heterostructure.

This work was motivated by experimental measurements $\Voc(\varphi)$ in superconductor/ferromagnet spin-valve structures by the same authors (manuscript under preparation)~\cite{Gonzalez-Ruano.2022}.
We also note that the possibility of thermopower reversal for antiparallel $\bm{m}$ and $\bm{m}'$ was mentioned briefly in Ref.~\cite{Ozaeta.2014}.
However, that paper does not elaborate on this as a potential control mechanism for the thermopower, discuss what junction parameters are required to observe a thermopower reversal, or calculate the angular dependence of the thermopower $S(\varphi)$.
As we demonstrate in this work, the angular dependence $S(\varphi)$ is highly non-trivial outside of the linear response regime.

Recently, a related \emph{bipolar thermoelectric effect} was also demonstrated experimentally by \citet{Germanese.2022}.
The fundamental mechanism in their setup is however different from our proposal:
Their thermopower is a multivalued function of the temperature difference due to a \emph{spontaneous} symmetry breaking between electrons and holes, and the sign of the thermopower is determined by the junction's bias history.
In our setup, the thermoelectric polarity is controlled by a separate magnetic input signal, and the thermopower is uniquely determined for each magnetic configuration.

Most previous studies employ the linear response approach, which is valid for very small temperature differences~$\Delta T$.
The electric current~$I$ is then approximated as a linear function of the voltage drop~$\Delta V$ and temperature difference~$\Delta T$,
\begin{equation}
	I \approx -G(\Delta V + S \Delta T),
\end{equation}
where $G$ is the conductance and $S$ the thermopower.
If the two ends of the device are short-circuited, there can be no net voltage drop across the device~$(\Delta V = 0)$, so a \emph{short-circuit current} $\Isc = -GS\Delta T$ must flow through the junction.
On the other hand, if the device is not part of a closed circuit, no electric current can flow through the device~$(I = 0)$.
Thus, an \emph{open-circuit voltage} $\Voc = -S\Delta T$ must form over the junction.
Since both $G$ and $S$ are constants in the linear response formalism, these observables are proportional: $\Isc = G \Voc$.

We here determine the full \emph{nonlinear response} of the system by numerically solving the nonequilibrium Usadel equation.
This enables us to study temperature differences~$\Delta T$ up to the the superconducting critical temperature~$\Tc$, which is more interesting with respect to applications.
The nonlinear response differs from the linear response in several ways.
As we will see in \cref{sec:results}, the observables $\Isc$ and $\Voc$ are no longer linearly related and in some cases exhibit a surprising angular dependence.
Furthermore, they have a highly nonlinear dependence on~$\Delta T$:
There is a near-quadratic response at low~$\Delta T$, which saturates and starts to decrease at moderate~$\Delta T$, and is even reversed at high~$\Delta T$.
This nonmonotonicity cannot be explained using linear response theory, which by definition predicts a purely linear relationship $\Voc \sim \Delta T$.

The rest of this paper is organized as follows.
\Cref{sec:method} describes how our numerical calculations were performed, including the material parameters and approximations used.
Readers that are most interested in the physical content and not the technical details can safely skip this section.
\Cref{sec:results} presents rigorously calculated physical predictions as well as a simple ``cartoon picture'' explanation of the underlying physical mechanism.
This section forms the core of this paper.
Finally, \cref{sec:conclusion} provides a conclusion and outlook.

\section{Methodology}\label{sec:method}
All results presented herein were obtained using the quasiclassical theory of superconductivity.
More specifically, we employ the Usadel equation \cite{Bergeret.2018, Chandrasekhar.2008, Belzig.1999, Rammer.1986, Usadel.1970}, which is valid for diffusive materials in and out of equilibrium.
Formally, our calculations presume a hierarchy of scales $\lambda \ll \ell_{\mathrm{e}} \ll \{ \xi, L \} < \{ \ell_{\mathrm{in}}, \ell_{\mathrm{sf}}, \ell_{\mathrm{so}} \}$, where $\lambda$ is the Fermi wavelength, $\ell_{\mathrm{e}}$ the elastic mean free path, $\xi$ the superconducting coherence length, $L$ the length of the FM, $\ell_{\mathrm{in}}$ the inelastic scattering length, and $\ell_{\mathrm{sf}}$ and $\ell_{\mathrm{so}}$ the spin-flip and spin-orbit scattering lengths.
If the Fermi wavelength and mean free path are not the two shortest length scales in the problem, then the Usadel equation is not formally valid.
If the inelastic scattering length and spin-dependent scattering lengths are not sufficiently long, then these mechanisms should be explicitly added to the Usadel equation.
However, we expect our predictions to remain qualitatively correct outside this parameter range as the ``cartoon pictures'' in \cref{sec:results} do not use the quasiclassical and diffusive approximations.
We note that spin-flip scattering has been shown to in some cases enhance the giant thermoelectric effect \cite{Rezaei.2018}.

The Usadel equation can be written in terms of a quasiclassical propagator~$\V{g}$, matrix current~$\V{\B{I}}$, and energy matrix~$\V{\Sigma}$,
\begin{equation}
	\begin{aligned}
		\nabla \cdot \V{\B{I}} &= i[\V{\Sigma}, \V{g}], &
		\V{\B{I}} &= -D\V{g}\nabla\V{g},
	\end{aligned}
	\label{eq:usadel}
\end{equation}
where $D$ is the diffusion coefficient.
Generally, the propagator contains local physical observables such as the density of states, while the matrix current contains transport properties such as the charge and heat currents.
All the matrices above have an $8\times8$ structure in  Keldysh$\otimes$Nambu$\otimes$Spin space,
\begin{equation}
	\begin{aligned}
		\V{g} &= \begin{pmatrix}
			\gr & \gk \\
			0\; & \ga
		\end{pmatrix}, &
		\V{\B{I}} &= \begin{pmatrix}
			\Ir & \Ik \\
			0\; & \Ia
		\end{pmatrix}, &
		\V{\Sigma} &= \begin{pmatrix}
			\H{\Sigma} & 0 \\
			0 & \H{\Sigma}
		\end{pmatrix},
	\end{aligned}
	\label{eq:keldysh}
\end{equation}
where each submatrix in the expansion above is left with a $4\times4$ structure in Nambu$\otimes$Spin space.
The remaining electron--hole and spin degrees of freedom are then described using $4\times4$ matrices $\H{\tau}_n$ and $\H{\sigma}_m$, respectively.
In terms of the usual $2\times2$ Pauli matrices $\{ \rho_0, \ldots, \rho_3 \}$, the basis matrices used above are $\forall n, m: \H{\tau}_n = \rho_n \otimes \rho_0,\, \H{\sigma}_m = \mathrm{diag}(\rho_m, \rho_m^*)$.
We also make use of the $4\times4$ Pauli vector $\B{\H{\sigma}} = (\H{\sigma}_1, \H{\sigma}_2, \H{\sigma}_3)$.
The energy matrix~$\H{\Sigma}$ describes the effective energies of quasiparticles, and its form depends on the particular materials under study.
To model \cref{fig:setup}, we only need to solve the Usadel equation inside the FM, in which case $\H{\Sigma} = \epsilon\H{\tau}_3 + \B{m}'\! \cdot \H{\B{\sigma}}$ where $\epsilon$ is the quasiparticle energy and $\bm{m}'$ is the magnetic exchange field.

To solve \cref{eq:usadel}, we also require boundary conditions that connect the solutions for $\V{g}$ inside the different metallic regions of \cref{fig:setup}.
For this purpose, we use boundary conditions that are valid for magnetic interfaces with low transparency and arbitrary spin polarization \cite{Ouassou.2017, Eschrig.2015lmf, Machon.2013, Cottet.2009, Cottet.2007}.
In terms of the matrix current~$\V{I} = \V{\B{I}} \cdot \B{n}$ that flows out of an interface with normal vector~$\B{n}$, this boundary condition can be written~\cite{Ouassou.2017}
\begin{align}
	(2L/D)\, \V{I} = (G_\textsc{t}/G_\textsc{n}) [\V{g}, F(\V{g}')] - i(G_\varphi/G_\textsc{n})[\V{g}, \V{m}], \label{eq:boundary} \\
	F(\V{v}) = \V{v} + \frac{P}{1+\sqrt{1-P^2}} \{ \V{v}, \V{m} \} + \frac{1 - \sqrt{1-P^2}}{1 + \sqrt{1-P^2}} \V{m} \V{v} \V{m}.
\end{align}
Here, $G_\textsc{n}$ is the Drude conductance of the material where we evaluate the boundary conditions, while $G_\textsc{t}$ and $G_\varphi$ are the tunneling and spin-mixing conductances of the interface.
The propagators $\V{g}$ and $\V{g}'$ describe the states at ``this side'' and ``the other side'' of the interface, respectively.
The interface magnetization enters via $\V{m} = \mathrm{diag}(\bm{m} \cdot \H{\B{\sigma}}, \bm{m} \cdot \H{\B{\sigma}})$. The spin polarization of the interface is~$P \in [0, 1]$.
Finally, $L$ and $D$ are the thickness and diffusion coefficient of the material where we evaluate the boundary conditions.
Here, we only solve \cref{eq:usadel} in the FM, so these parameters describe that layer.

The solution of \cref{eq:usadel,eq:boundary} can be made more tractable by exploiting the symmetries of the quasiclassical propagator~$\V{g}$.
It can be shown that the retarded and advanced propagators are related by an electron--hole symmetry $\ga = {-\H{\tau}_3 \H{g}^{\textsc{r}\dagger} \H{\tau}_3}$, while the Keldysh propagator is related to the nonequilibrium distribution function~$\H{h}$ via $\gk = \gr \H{h} - \H{h} \ga$.
Thus, it is in practice sufficient to calculate $\gr$ and $\H{h}$.
Let us first discuss $\gr$.
This propagator satisfies a normalization condition $(\H{g}^\textsc{r})^2 = \H{\tau}_0$ and electron--hole symmetry $\H{g}^\textsc{r}(-\epsilon) = -\H{\tau}_1 \H{g}^{\textsc{r}*}(+\epsilon) \H{\tau}_1$.
We account for these symmetries via the Riccati parametrization~\cite{Jacobsen.2015, Schopohl.1998},
\begin{equation}
	\H{g}^{\textsc{r}} = 
	\begin{pmatrix}
		+N & 0 \\ 0 & -\tilde{N}
	\end{pmatrix}
	\begin{pmatrix}
		1 + \gamma\tilde{\gamma} & 2\gamma \\
		2\tilde{\gamma} & 1 + \tilde{\gamma}\gamma
	\end{pmatrix},
\end{equation}
where the normalization matrix is defined as $N \equiv (1 - \gamma\tilde\gamma)^{-1}$ and the electron--hole conjugation $\tilde\gamma(+\epsilon) \equiv \gamma^*(-\epsilon)$.
This parametrization is well-suited for numerical computation since the $2\times2$ matrices $\gamma, \tilde\gamma$ are single-valued and bounded.
Next, we discuss the distribution function~$\H{h}$.
This matrix can be taken as block-diagonal~\cite{Chandrasekhar.2008} and satisfies an electron--hole symmetry $\H{h}(-\epsilon) = -\H{\tau}_1 \H{h}^*(+\epsilon) \H{\tau}_1$.
We parametrize it as \cite{Ouassou.2018fbw, Bergeret.2018, Chandrasekhar.2008, Belzig.1999}
\begin{equation}
	\H{h} = \sum_{nm} h_{nm} \H{\tau}_n \H{\sigma}_m,
\end{equation}
where the sum is taken over indices ${m \in \{0, 1, 2, 3\}}$ and ${n \in \{0, 3\}}$ that yield block-diagonal matrices.
The expansion coefficients $h_{nm} \equiv (1/4) \mathrm{Tr}[\H{\tau}_n \H{\sigma}_m \H{h}]$ are real-valued.

By applying the parametrizations above to \cref{eq:usadel,eq:keldysh}, one can derive a second-order nonlinear differential equation for the Riccati parameters~$\gamma, \tilde\gamma$ and a second-order linear differential equation for the distribution traces~$h_{nm}$.
These derivations are described in detail in \eg Ref.~\cite{Jacobsen.2015} and Ref.~\cite{Ouassou.2018fbw}, respectively, so we do not repeat them here.
The resulting boundary value problems are then solved numerically using the open-source software package \href{https://github.com/jabirali/geneus}{github.com/jabirali/geneus}, which employs the numerical procedures described in Ref.~\cite{Ouassou.Thesis}.

Once we have a solution, the electric current~$\B{I}$ is extracted from the calculated Keldysh matrix current~$\H{\B{I}}^\textsc{k}$ via~\cite{Ouassou.201984c, Bergeret.2018} 
\begin{equation}
	\bm{I} = -\frac14 eN_0 \int\limits_0^\infty \mathrm{d}\epsilon\, \mathrm{Re}\,\mathrm{Tr}\big\{\H{\tau}_3\H{\sigma}_0\H{\B{I}}^\textsc{k}(\epsilon)\big\},
	\label{eq:current}
\end{equation}
where $e < 0$ is the electron charge and $N_0$ is the density of states at the Fermi level in the absence of superconductivity.

\begin{figure}[b!]
	\includegraphics[width=\columnwidth]{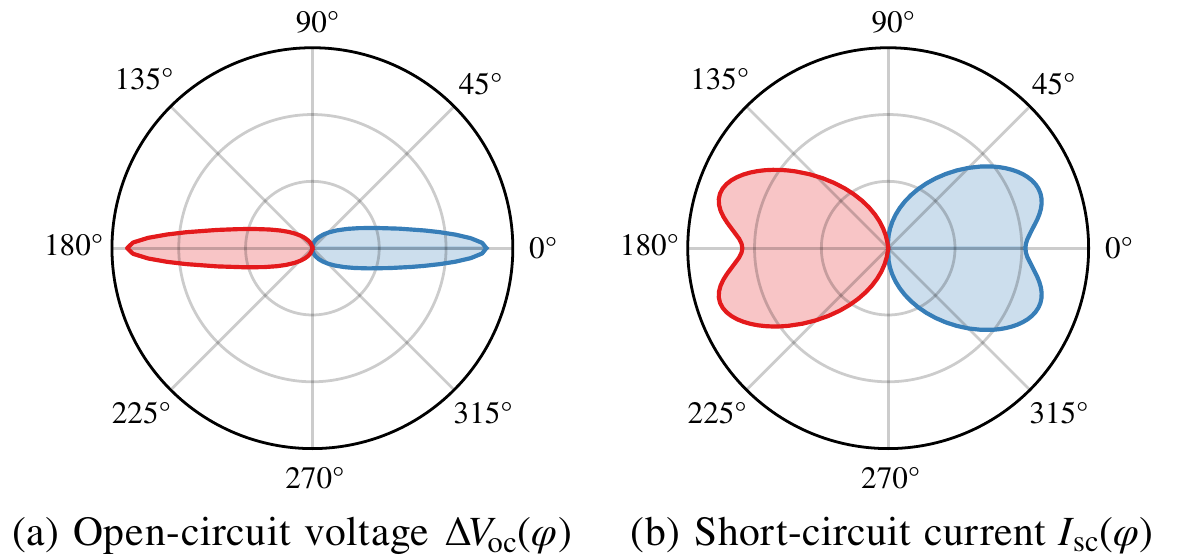}
	\caption{
		Numerical results for the experiment proposed in \cref{fig:setup}.
		The polar angle is the magnetic misalignment~$\varphi$ between the two ferro\-magnets, while the radius corresponds to $|\Voc|$~and~$|\Isc|$, respectively.
		The color shows the thermoelectric polarity, \ie whether the predicted values for $\Voc$ and $\Isc$ are positive (blue) or negative (red).
		The total radius shown is (a)~$\num{1.1e-2}\, V_0$ and (b)~$\num{6.5e-5}\,I_0$.
	}
	\label{fig:polarheat}
\end{figure}

Let us now discuss the specific material parameters used to model the setup in \cref{fig:setup}.
The S was treated as a reservoir, so we used the propagator $\H{g}^\textsc{r}_\textsc{s}$ of a bulk superconductor and the equilibrium distribution $\H{h}_\textsc{s} = \tanh(\epsilon/2T_\textsc{s})\,\H{\tau}_0$.
The superconducting gap $\Delta \approx \Delta_0 \tanh\!\big[ 1.74 \sqrt{\Tc/T_\textsc{s} - 1} \big]$, where $\Delta_0$ is the zero-temperature bulk gap.
The FM has length $L = \xi/2$ and exchange field $\bm{m}' = m' (\cos\varphi\,\bm{e}_x - \sin\varphi\,\bm{e}_y)$, where $\xi = \sqrt{D/\Delta_0}$ is the coherence length in~S.
The N was treated as a voltage-biased normal-metal reservoir, so we used the bulk propagator ${\H{g}^\textsc{r} = \H{\tau}_3}$ and nonequilibrium distribution $\H{h} = \mathrm{diag}(h_+, h_+, h_-, h_-)$ where $h_\pm = \tanh[(\epsilon \pm e \Delta V)/2T_\textsc{n}]$.
For the main simulations in \cref{fig:polarheat}, we used $T_\textsc{s} = \Tc/100$, $T_\textsc{n} = \Tc/3$, and $m' = 3\Delta_0$, although these default parameters were varied in the parameter study that follows.

We now discuss the parameters of the boundary conditions.
Since the FM is assumed to be a weak ferromagnet, we neglect the magnetic properties $\Gphi$ and $P$ of the S/FM interface.
The remaining interface parameter is the tunneling conductance, which we set to a moderate value $\GT = 0.3\GD$ where $\GD$ is the Drude conductance of the~FM.
Note that the inclusion of $\Gphi$ would not qualitatively change the results, since it provides the same physical effects as adjusting the exchange field~$\bm{m}'$ in the FM layer.
Introducing a \emph{large} spin polarization~$P$ would however change the thermopower~$S(\varphi)$, since the FI would no longer dominate the spin filtering in the junction.
The FM/FI/N boundary was modeled as a magnetic interface, where quasiparticles effectively tunnel through a spin-dependent barrier between the FM and N layers.
Since this barrier is insulating, we selected a low conductance $\GT = 0.1\GD$.
Moreover, we take the FI to be a strong ferromagnet, and assume a complete spin polarization~$P = 1$, moderately high spin-mixing conductance $\Gphi = 1.25$, and fixed magnetization direction $\bm{m} = \bm{e}_x$.
We note that the thermopower in the junction depends roughly linearly on~$P$, but is not very sensitive to $\Gphi$ since the FM dominates the spin splitting in the proposed device.

The model above has two free parameters: the magnetic misalignment angle $\varphi$ between the FM and FI, and the voltage difference~$\Delta V$ between the N and S.
We performed parallel simulations for various configurations of these parameters, and used \cref{eq:current} to obtain the magnetically-dependent current--voltage relation $I[\varphi, \Delta V]$.
The physical predictions were then extracted from this data.
The short-circuit current is simply the current at zero voltage bias $\Isc(\varphi) \equiv I[\varphi, 0]$.
The open-circuit voltage is the voltage that makes the current vanish, which had to be interpolated from $I[\varphi, \Voc(\varphi)] \equiv 0$.
These are the two experimental signatures that we focus on in this paper.

\section{Results and discussion}\label{sec:results}
Let us first summarize the unit system used below.
As base units for lengths, temperatures, and energies we take the superconducting coherence length~$\xi$, critical temperature~$\Tc$, and zero-temperature gap~$\Delta_0$.
Voltages are then naturally measured in terms of $V_0 \equiv \Delta_0/|e|$, where $e$ is the electron charge.
Electric currents are specified in the material-dependent unit $I_0 \equiv |e|AN_0 \Delta_0^2 \xi / \hbar$, where $A$ is the junction's cross-sectional area and $N_0$ is the Fermi-level density of states above~$\Tc$.
The last two parameters, as well as the diffusion coefficient~$D$, are assumed to be the same in every metallic region of \cref{fig:setup}.
The corresponding thermopower unit is $S_0 \equiv V_0/\Tc \approx \SI{152}{\upmu\volt\per\kelvin}$, which follows from the BCS ratio $\Delta_0/k_\textsc{b}\Tc \approx 1.764$.
Most results presented here are in the range 1--10~\% of~$S_0$; higher values can \eg be obtained if the FM is made shorter, the interfaces more transparent, or $\Delta T$ is increased.

\cref{fig:polarheat} shows the main numerical results for the magnetically controlled thermoelectric generator proposed in \cref{fig:setup}.
These results were obtained for a low temperature $T = \Tc/100$ in~S and moderate temperature $T = \Tc/3$ in~N, such that their temperature difference is $\Delta T \approx \Tc/3$.
This temperature difference then drives a thermoelectric response in the device, which can manifest as either an open-circuit voltage~$\Voc(\varphi)$ or short-circuit current~$\Isc(\varphi)$ depending on the switch in \cref{fig:setup}.
We see that the thermoelectric response depends sensitively on the magnetic misalignment angle~$\varphi$, which can be experimentally controlled by \eg rotating an applied magnetic field.

Previous studies considered parallel magnetizations (${\varphi = 0^\circ}$), where the thermoelectric signatures in \cref{fig:polarheat} are positive.
If one magnetization is rotated until the two become perpendicular (${\varphi = 90^\circ}$), thermoelectricity vanishes.
Further rotating until the magnets are antiparallel (${\varphi = 180^\circ}$), the thermoelectric effect is completely reversed: $\Voc(180^\circ) \approx -\Voc(0^\circ)$ and $\Isc(180^\circ) \approx -\Isc(0^\circ)$.
\emph{This demonstrates the complete magnetic control over the giant thermoelectric effect.}
These results can be interpreted as a magnetically-dependent thermopower~$S(\varphi)$ that varies from a large positive value to a large negative value.
However, note that in nonlinear response the thermopower also depends on temperature and voltage.

\begin{figure*}
	\includegraphics[width=\textwidth]{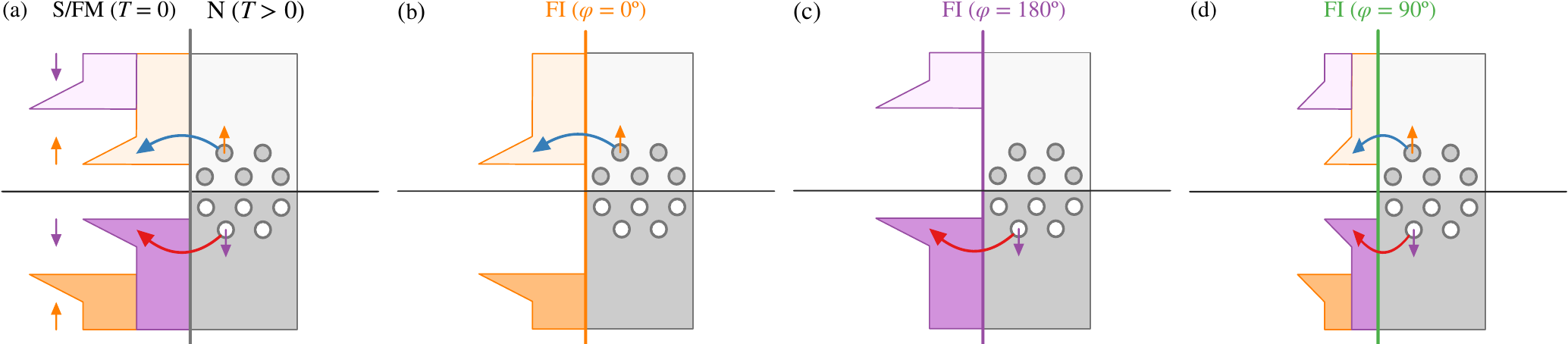}
	\caption{
		Cartoon picture of the mechanism behind the proposed effect.
		\textbf{(a)}~On the left we sketch the spin-split density of states in an S/FM bilayer at ${T = 0}$.
		All states below the Fermi level are filled and all states above it are empty.
		On the right is an N~layer at ${T > 0}$.
		Spin-up electrons and spin-down holes can tunnel into the lowest-energy states of the S/FM subsystem, creating a net heat and spin flow to the left.
		However, the electron flow (blue) and hole flow (red) have opposite charges and equal magnitude, so there is no net electric current.
		\textbf{(b)}~A~fully spin-up polarized~FI is inserted between the S/FM and N subsystems.
		The quasiparticles in~N can now only ``see'' the spin-up bands of the S/FM subsystem.
		Electrons can thus tunnel like before, but the lowest-energy hole states are no longer accessible, which strongly suppresses the hole flow.
		Thus, net-negative charge flows to the left due to the thermal excitations.
		\textbf{(c)}~The FI is now polarized in the spin-down direction.
		Since the excitations in~N only see the spin-down bands in S/FM, mainly holes can tunnel to the left, and we get a net-positive charge flow.
		\textbf{(d)}~When the FI is polarized perpendicularly to the FM, half of each spin band is visible.
		The resulting electric currents cancel as in panel~(a).
	}
	\label{fig:mechanism}
\end{figure*}

\begin{figure*}
	\includegraphics[width=\textwidth]{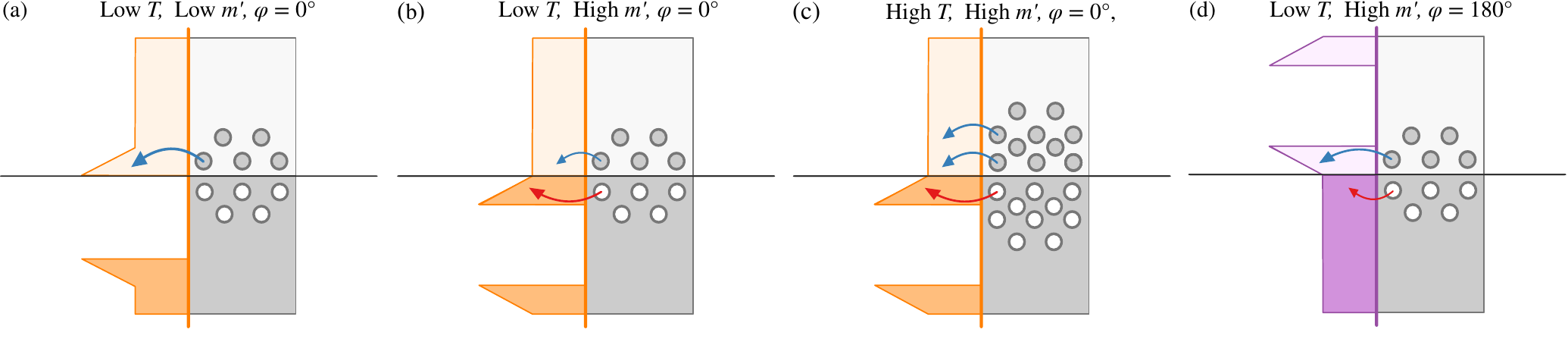}
	\caption{
		Cartoon picture explaining the sign reversal of the thermoelectric effect.
		\textbf{(a)}~Let us first consider $\varphi = 0^\circ$ as in \cref{fig:mechanism}(b).
		As long as the spin-splitting field $m' \lesssim \Delta_0$, there is a low-energy coherence peak in the electron band and a gap in the hole band, resulting in an electron-dominated thermoelectric effect.
		\textbf{(b)}~If $m' \gtrsim \Delta_0$, the spin splitting is sufficiently large to push the coherence peak down into the hole band. 
		This is an oversimplification of the actual spin-dependent density of states in such a junction, but the explanation is qualitatively correct and yields the correct physical intuition.
		In this case, both electrons and holes can tunnel from the hot N layer, but the coherence peak makes the hole current dominant at low temperatures.
		This reverses the sign of the thermopower compared to panel~(a).
		\textbf{(c)}~If the temperature of the normal metal is increased, tunneling contributions at moderate energies become increasingly important.
		This increases the electron current more than the hole current due to the gap in the hole band, and at sufficiently high temperatures the sign of the thermopower therefore flips again.
		The effect is exacerbated if the S is heated as well, since more of the states in the coherence peak will then already be occupied, which again increases the importance of tunneling processes at higher energies.
		\textbf{(d)}~This shows the equivalent to panel~(b) for $\varphi=180^\circ$.
		The thermopower $S(\varphi)$ remains antisymmetric for $m' \gtrsim \Delta_0$ since the spin-up and spin-down bands are still shifted equally in opposite directions.
	}
	\label{fig:mechanism2}
\end{figure*}

The main mechanism behind the magnetically-dependent thermoelectric effect is sketched in \cref{fig:mechanism}.
The crux of the matter is that a spin-splitting field~$\bm{m}'$ gives rise to a spin-dependent electron--hole asymmetry in a superconducting system~\cite{Ozaeta.2014}.
We can tap into this asymmetry via a spin-filtering field~$\bm{m}$, which determines whether an adjacent metal couples predominantly to the electrons or holes of the system above.
The dominant energy carriers can then be either electrons or holes depending on the misalignment angle~$\varphi$ between $\bm{m}$ and~$\bm{m}'$.
When a heat gradient is applied, the resulting thermoelectric effect depends on the charge of the dominant energy carrier.
This explains how the magnetic misalignment angle~$\varphi$ can modulate the sign of the giant thermoelectric effect.

\begin{figure*}
	\includegraphics[width=\textwidth]{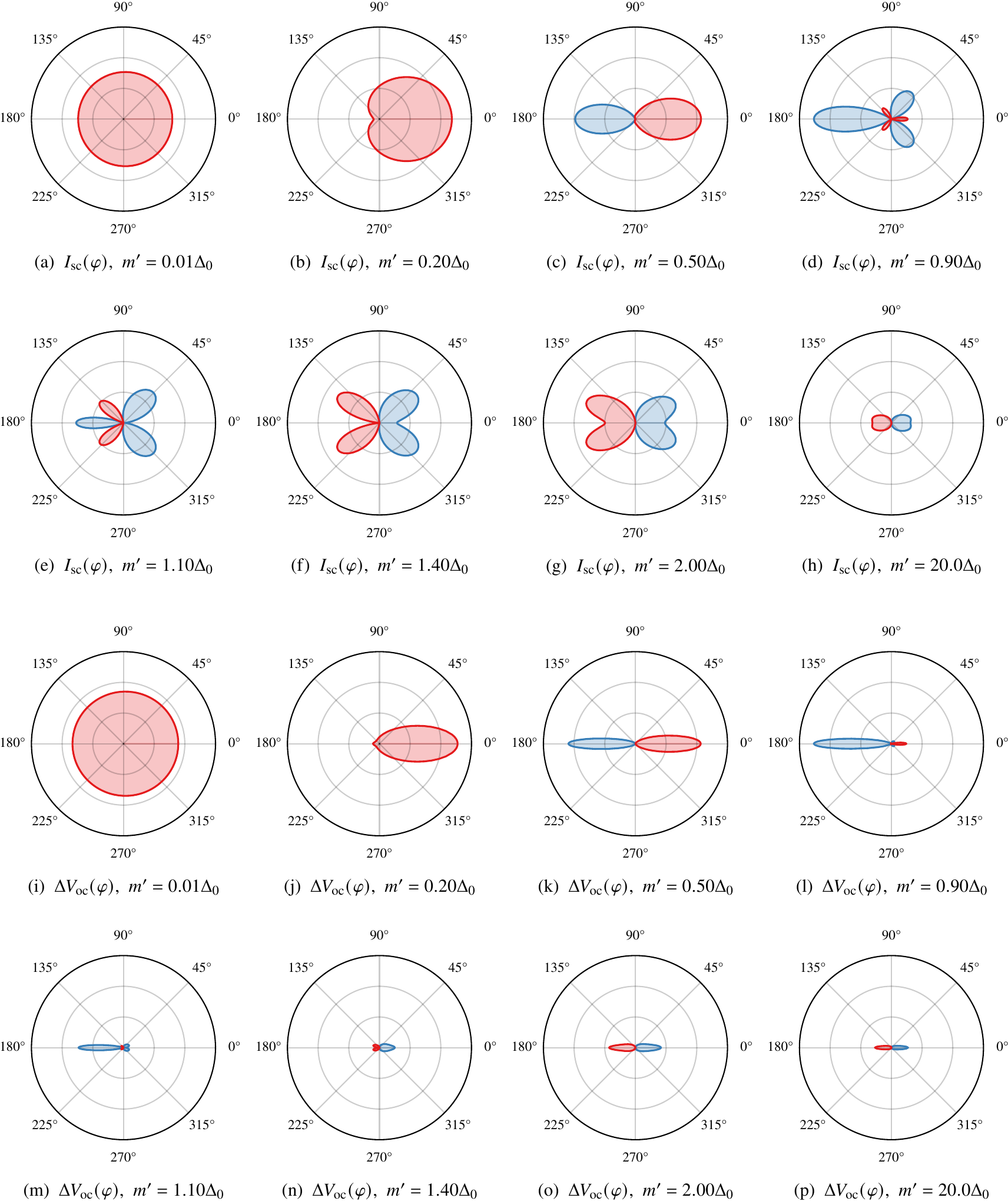}
	\caption{
		Evolution of the thermoelectric response with the strength of the magnetic exchange field $m'$ in the FM.
		Panels (a--h) show the short-circuit current $\Isc(\varphi)$, where the plot radius is $1.2\times10^{-4}\,I_0$.
		Panels (i--p) show the open-circuit voltage $\Voc(\varphi)$, where the plot radius is $3\times10^{-2}\,V_0$.
		Except for the exchange field and radial scale, all parameters used for the simulations and visualization are the same as in \cref{fig:polarheat}.
	}
	\label{fig:details}
\end{figure*}

The results above were obtained for an FM with moderate magnetic exchange field $m' = 3\Delta_0$.
However, as shown in \cref{fig:details}, the predictions for $\Isc(\varphi)$ and $\Voc(\varphi)$ change drastically for weaker ferromagnets.
In the limit $m' \rightarrow 0$, both $\Isc(\varphi)$ and $\Voc(\varphi)$ become independent of $\varphi$.
We refer to this rotational invariance as a \emph{symmetric} thermoelectric effect.
For slightly stronger ferromagnets $m' = 0.20\Delta_0$, we find an \emph{asymmetric} thermoelectric effect:
$\varphi$ acts as a magnetic on--off switch and tunes the system between zero and maximum thermoelectric response.
Further increasing the field to $m' = 0.50\Delta_0$, we obtain an \emph{antisymmetric} thermoelectric effect.
Here, $\varphi$ tunes the thermopower between positive and negative values of similar magnitude.
The symmetric, asymmetric, and antisymmetric scenarios above are well-described by the angular dependencies $\Isc(\varphi) \sim 1,\, \Isc(\varphi) \sim 1 + \cos\varphi,\, \Isc(\varphi) \sim \cos\varphi$, while the voltage curves $\Voc(\varphi)$ are more pronounced for parallel and antiparallel orientations of the two magnets.

This transition can be understood as follows.
When ${m' \rightarrow 0}$, the FM becomes a normal metal and thus $\varphi$ loses all physical significance.
There is still a thermoelectric effect because the FI itself can provide the necessary spin splitting, but this spin splitting is always along the same magnetic axis as the spin filtering.
This produces a $\varphi$-independent thermoelectric effect.
As $m' \rightarrow \Delta_0/2$, the FM exchange field becomes strong enough to dominate the spin splitting of the superconducting density of states.
In this regime, the mechanism in \cref{fig:mechanism} explains why an antisymmetric thermoelectric response ensues.
It is perhaps not surprising that for intermediate fields $m' = 0.20\Delta_0$, the thermoelectric response $\Isc(\varphi) \sim 1 + \cos\varphi$ looks like an average between the responses at $m' \rightarrow 0$ and $m' \rightarrow \Delta_0/2$.
Physically, we can interpret this asymmetric response as a point where the spin-splitting effects of the FI and FM are comparable in size, such that they either add up or cancel depending on the relative orientations of the two magnets.
Note that the values of $m'$ for which these symmetry transitions occur depend on junction parameters such as \eg the FM length~$L$.

For stronger ferromagnets $m' \gtrsim 2\Delta_0$, the results in \cref{fig:details} again show an antisymmetric thermoelectricity~$S(\varphi)$.
However, the sign of the thermopower is flipped compared to $m' \leq \Delta_0/2$; the reason for this sign reversal is explained in \cref{fig:mechanism2}.
Moreover, there is a striking deviation between the shapes of $\Isc(\varphi)$ on one hand, which peaks along $0\pm45^\circ$ and $180\pm45^\circ$; and $\Voc(\varphi)$ on the other hand, which peaks along $0^\circ$ and $180^\circ$.
Numerically, we find that the magnetic configurations that maximize $\Isc(\varphi)$ are the same that produce non-negligible supercurrent contributions (not shown).
Since the generation of equal-spin triplet Cooper pairs is maximized for $\varphi = \pm 90^\circ$ while the giant thermoelectric effect is maximized for $\varphi = 0^\circ, 180^\circ$, the largest supercurrents are naturally found at the intermediate angles where both effects are significant: the ``diagonals'' $0\pm45^\circ$ and $180\pm45^\circ$.
These supercurrents however flow in the \emph{opposite direction} from the resistive thermoelectric currents;
this phenomenon is known from the study of \eg Josephson junctions as a thermophase effect~\cite{Giazotto.2015z7}.
Thus, the supercurrents themselves are not the reason why the currents along the $45^\circ$ diagonals are large, but rather a signature that these misalignments maximize the triplet superconductivity in FM which in turn enhances $\Isc(\varphi)$.
The explanation for this enhancement is most likely related to how odd-frequency superconductivity increases the density of states in the energy ranges corresponding to such correlations.
For very large fields $m' \gg \Delta_0$, thermoelectricity stabilizes in an antisymmetric shape that becomes insensitive to the exact value of~$m'$, and $\Isc(\varphi)$ becomes more similar to $\Voc(\varphi)$.
This is because large $m'$ destroys superconducting correlations before they can be converted into robust equal-spin triplets.

Finally, we discuss the intermediate regime $m' \approx \Delta_0$.
\cref{fig:mechanism2} shows that the system transitions between a range of exotic shapes for $\Isc(\varphi)$ in this regime.
As discussed above, significant supercurrent contributions are found along the ``diagonals'' $0\pm45^\circ$ and $180\pm45^\circ$, and we find that $m' \approx \Delta_0$ maximizes these supercurrents for our junction parameters.
Moreover, a detailed analysis reveals a significant spin-valve-like effect, whereby $\varphi$ tunes the density of states between regimes with a zero-energy plateau, zero-energy peak, both, and neither. 
This modulation complicates the mechanism in \cref{fig:mechanism}: the spin-split density of states, equal-spin triplet generation, and spin-dependent tunneling all depend sensitively on~$\varphi$, which makes it difficult to predict the angular dependence $S(\varphi)$ without explicit calculation.
This exotic angular dependence is technically present in both $\Voc(\varphi)$ and $\Isc(\varphi)$, but the effect is so much larger in magnitude for $\Isc(\varphi)$ that it might be difficult to observe if only $\Voc(\varphi)$ is measured experimentally.

\begin{figure}[t!]
	\includegraphics[width=\columnwidth]{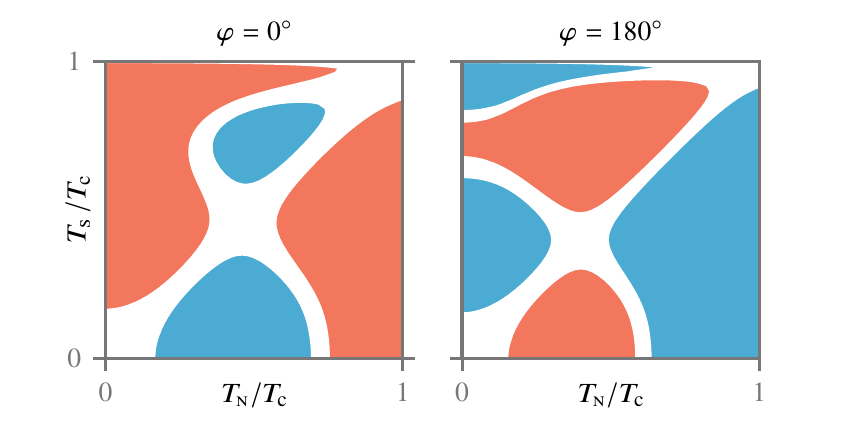}
	\caption{
		Thermoelectric current~$\Isc(\varphi)$ as function of the temperatures $T_\textsc{s}$ in S and $T_\textsc{n}$ in~N for $m'=3\Delta_0$.
		Blue regions correspond to ${\Isc > +I_\text{min}}$ and red regions to ${\Isc < -I_\text{min}}$, where $I_\text{min} \equiv \num{e-5}~I_0$.
	}
	\label{fig:temperature-params}
\end{figure}

We now return to the main results in \cref{fig:polarheat}, which correspond to a ferromagnet with $m' = 3\Delta_0$.
\cref{fig:temperature-params} shows the temperature dependence of the effect for this junction.
When $T_\textsc{s} \lesssim \Tc/2$, the antisymmetry between $\varphi = 0^\circ$ and $\varphi = 180^\circ$ is nearly perfect, demonstrating the robustness of the magnetically controlled thermoelectricity over a large range of temperatures.
However, this symmetry is broken when $T_\textsc{s} \rightarrow \Tc$.
This is explained by a spin-valve-like effect:
Since the gap~$\Delta$ in S decreases in this region, the proximity-induced minigap~$\Delta'$ in FM must also decrease.
This makes the minigap more vulnerable to the pair-breaking effects of ferromagnetism, which is stronger when FM and FI are oriented in the same direction.
Naturally, $\Isc$ is reversed once one crosses the line $T_\textsc{s} = T_\textsc{n}$, since $\Delta T$ then points in the opposite direction.
Interestingly, we also observe a reversal once the temperature in the FM $T_\textsc{fm} \approx (T_\textsc{s} + T_\textsc{n})/2$ reaches a threshold $\sim\! \Tc/2$.
{The fact that large temperatures should reverse the thermopower when $m' > \Delta_0$ is can be understood from the sketches in \cref{fig:mechanism2}.}\\

\section{Conclusion and outlook}\label{sec:conclusion}
We have established that the giant thermoelectric effect in superconducting devices can be completely controlled via a magnetic control knob.
We provided a simple explanation of the physical mechanism, a concrete proposal for experimental realization, and quantitative predictions beyond the usual approximation of linear response.
Moreover, we demonstrated how the angular dependence of the thermopower evolves between symmetric, asymmetric, and antisymmetric shapes, including some highly non-trivial and unexpected intermediate shapes. Finally, we highlighted how the thermoelectric effect varies non-monotonically with \eg the temperature.

Our findings also point towards interesting avenues for further research.
While we focus on the magnetically controlled Seebeck effect herein, the Onsager relations imply the existence of a related Peltier effect.
Thus, the same device can likely act as a magnetically tunable ``heat pump''.
Controlling the sign of the thermopower (analogous to the inversion of thermoelectric signals between $p$- and $n$-doped semiconductors) could be crucial for the design of Peltier elements based on superconducting spin valves.
Another possibility is to replace our N with another S to produce a Josephson junction.
Such a device will likely exhibit a magnetically tunable thermophase.

\begin{acknowledgments}
	The authors acknowledge Coriolan Tiusan and Michel Hehn.
	J.A.O. and J.L. were supported by the Research Council of Norway through Grant No. 323766, and through its Centres of Excellence funding scheme Grant No. 262633 ``QuSpin.''
	The work in Madrid was supported by Spanish Ministerio de Ciencia (RTI2018-095303-B-C55, PID2021-124585NB-C32) and Consejer\'ia de Educaci\'on e Investigaci\'on de la Comunidad de Madrid (NANOMAGCOST-CM Ref. P2018/NMT-4321) Grants.
	FGA acknowledges financial support from the Spanish Ministry of Science and Innovation, through the ``Mar\'ia de Maeztu'' Program for Units of Excellence in $R\&D$ (CEX2018-000805-M) and ``Acci\'on financiada por la Comunidad de Madrid en el marco del convenio plurianual con la Universidad Aut\'onoma de Madrid en L\'inea 3: Excelencia para el Profesorado Universitario''.
\end{acknowledgments}

% Original bib.
% \clearpage
% \bibliography{references}

% Compiled bbl.
%
\end{document}